\documentclass[%
 reprint,
 amsmath,amssymb,
 aps,
 onecolumn,
 notitlepage
]{revtex4-1}

\usepackage{graphicx}
\usepackage{dcolumn}
\usepackage{bm}

\usepackage{braket}
\usepackage{listings}
\usepackage{color}
\usepackage{enumitem}   

\definecolor{codegreen}{rgb}{0,0.6,0}
\definecolor{codegray}{rgb}{0.5,0.5,0.5}
\definecolor{codepurple}{rgb}{0.58,0,0.82}
\definecolor{backcolour}{rgb}{0.95,0.95,0.92}

\def\note#1{}

\lstdefinestyle{mystyle}{
    backgroundcolor=\color{backcolour},
    commentstyle=\color{codegreen},
    keywordstyle=\color{magenta},
    numberstyle=\tiny\color{codegray},
    stringstyle=\color{codepurple},
    basicstyle=\footnotesize,
    breakatwhitespace=false,
    breaklines=true,
    captionpos=b,
    keepspaces=true,
    numbers=left,
    numbersep=5pt,
    showspaces=false,
    showstringspaces=false,
    showtabs=false,
    tabsize=2
}

\begin{document}


\title{Repeat-Until-Success circuits with fixed-point oblivious amplitude amplification}

\author{Gian Giacomo Guerreschi}
 \email{gian.giacomo.guerreschi@intel.com}
\affiliation{Intel Labs}


\date{\today}


\begin{abstract}
Certain quantum operations can be built more efficiently through a procedure known as Repeat-Until-Success. Differently from other non-deterministic quantum operations, this procedure provides a classical flag which certifies the success or failure of the procedure and, in the latter case, a recovery step allows the restoration of the quantum state to its original condition. The procedure can then be repeated until success is achieved. After success is certified, the RUS procedure can be equated to a coherent gate. However, this is not the case when the operation needs to be conditioned on the state of other qubits, possibly being in a superposition state. In this situation, the final operation depends on the failure and success history and introduces a ``distortion'' that, even after the final success, depends on the past outcomes.
We quantify the distortion and show that it can be reduced by increasing the probability of success towards unity. While this can be achieved via oblivious amplitude amplification when the original success probability is known, we propose the use of fixed-point oblivious amplitude amplification to reduce this unwanted distortions below any given threshold even without knowing the initial success probability.
\end{abstract}

\pacs{Valid PACS appear here}

\maketitle



\section{Introduction}
\label{sec:introduction}

Not all measurement processes modify quantum information in irreversible ways. In certain carefully constructed circumstances, the knowledge of the outcome value is sufficient to reverse the effect of the measurement. Quantum error correcting codes are built on this observation: error syndromes are measured to provide insight on the errors that have occurred without compromising the quantum information, but rather allowing for its perfect recovery. This is also the case for the class of Repeat-Until-Success (RUS) quantum circuits \cite{Wiebe2013,Paetznick2014}, in which ancilla qubits are entangled with the data qubits and then measured. The measurement outcome determines whether the desired operation was successfully implemented on the data qubits or if another, but reversible, operation was alternatively performed. In case of failure, one can easily correct the unwanted result of the previous attempt and try again. As indicated by its name , one can repeat the procedure until success is achieved and certified by the correct measurement outcome. Notice that the overall failure probability decreases exponentially with the number of attempts, therefore a small number of repetitions usually suffice.

RUS circuits represent non-deterministic constructions, but the overall operation can be considered deterministic since one has access to a classical flag, \emph{i.e.} the ancilla measurement outcome, that certifies the correct implementation. No exponential cost needs to be paid when multiple RUS operations are present in the same quantum circuit, as long as they are not nested. The overhead is essentially a constant factor representing the average number of attempts before achieving success for a single RUS operation.

RUS circuits have so far found application in the synthesis of arbitrary rotations using discrete gate sets \cite{Paetznick2014,Bocharov2015} and in the recent proposal of a quantum neuron designed as a modular unit for quantum neural networks \cite{Cao2017a}. However, the latter application revealed that RUS constructions cannot be considered equivalent to traditional unitary gates since they cannot be performed in a (quantum) conditional way without introducing distortions.

In this Article, we explain this effect, quantify the corresponding amplitude distortion and present a few approaches to mitigate it. A broadly applicable solution is provided by the application of fixed-point oblivious amplitude amplification since the distortion vanishes when the probability of success of the RUS circuit tends to unity. In addition, we quantify and compare the cost in terms of T gates of several oblivious amplitude amplification (OAA) techniques.

The content of this work is organized as follows: section~\ref{sec:rus-operations} describe the RUS construction and previous results of applying oblivious amplitude amplification; section~\ref{sec:controlled-rus} introduces the problem of amplitude distortion for conditional implementations and quantifies the undesired effect; section~\ref{sec:deterministic} presents a solution in terms of deterministic OAA viable when the success probability of the unconditional RUS circuit is known a priori; section~\ref{sec:fixed-point} proposes the application of fixed-point OAA to reduce the amplitude distortion below an arbitrary small threshold even without prior knowledge of the success probability. Finally we discuss the remaining limitations of conditional RUS circuits and draw conclusions.


\section{RUS operations}
\label{sec:rus-operations}

Repeat-Until-Success operations have been introduced in the context of gate decomposition, where one typically desires to implement an arbitrary single-qubit rotation by expressing it as a sequence of gates belonging to a specific (often discrete) set. Reference \cite{Paetznick2014} characterizes which operation can be represented in terms of RUS circuits and \cite{Bocharov2015} provides an efficient way to obtain such representation.

Consider a single-qubit RUS operation that requires $m$ ancilla qubits, the case of multi-qubit operations follow in a straightforward way. We describe this operation by a $(m+1)$ unitary operation $A$, followed by measurement in the computational basis for all $m$ ancillas and by active feedback: if the measurement outcome corresponds to state $\ket{00\dots0}$, then the operation $U$ has been successfully implemented on the data qubit; otherwise one of the unwanted (but reversible) operations $R_i$ has been implemented, and one needs to correct it before performing a new attempt. Formally:
\begin{equation}
\label{eq:rus-state}
    A \ket{0^m}\ket{\psi} = \sqrt{\lambda_0} \ket{0^m} U \ket{\psi}
            + \sum_{i=1}^{2^m-1} \sqrt{\lambda_i} \ket{i} R_i \ket{\psi} \; ,
\end{equation}
with $\lambda_0$ being the probability of success for implementing $U$, and $\sum_{i\neq0} \lambda_i = 1-\lambda_0$ due to the unitarity of A. Notice that the amplitude for $\ket{0^m}U\ket{\psi}$ is assumed to be real and positive, otherwise the opposite phase could be easily introduced in a modified version of $A$ to achieve the cancellation. Relative phases arising in the distinct failure cases have been absorbed in the definition of $R_i$ leaving all $\lambda_i$ real and non-negative.
To properly define RUS operations, the success probability $\lambda_0$ should not depend on $\ket{\psi}$. The number of repetitions required to achieve success is a stochastic variable distributed geometrically with expected value $\tfrac{1}{\lambda_0}$ and variance $\tfrac{(1-\lambda_0)}{\lambda_0^2}$.

As observed in \cite{Paetznick2014}, one can implement oblivious amplitude amplification (OAA) to increase the overlap between the final state and the desired state $\ket{0^m}U\ket{\psi}$. To accomplish this, one needs to perform the inverse circuit $A^\dagger$ together with the reflection
\begin{equation}
\label{eq:reflection}
    S_\pi = (I^m - 2 \ket{0^m}\bra{0^m}) \otimes I \; ,
\end{equation}
with $I^m$ denoting the identity operation on $m$ qubits. The notation $S_\pi$ has been chosen for consistency with a more general operation to be introduced in a later section. Depending on the number of repetitions $j$, the final state corresponds to (see reference \cite{Berry2014} for explicit derivation):
\begin{equation}
\label{eq:amplitude-amplification}
    (-A S_\pi A^\dagger S_\pi)^j A \ket{0^m}\ket{\psi} =
        \sin( (2j+1) \theta) \ket{0^m} U \ket{\psi}
        + \cos( (2j+1) \theta) \ket{\Phi^\perp} \; ,
\end{equation}
with $\sin(\theta)=\sqrt{\lambda_0}$ and $\ket{\Phi^\perp}$ being a $(m+1)$ qubit state orthogonal to $\ket{0^m}U\ket{\psi}$, and actually having no overlap with the support of $\ket{0^m}\bra{0^m} \otimes I$.

In the decomposition of quantum operations into a discrete, universal set of gates like $\{H,S,T,CNOT\}$, the cost of implementing $U$ is usually computed in terms of the number of $T$ gates alone, since fault-tolerant implementations of the $T$ gate are orders of magnitude more expensive than implementations of any Clifford gate \cite{Reiher2017}. Using $C_T(G)$ to indicate the number of T gates required to implement gate $G$, one has:
\begin{align*}
    \text{without amplitude amplification } &\rightarrow\quad C_T(U) = \Big|\sin(\theta)\Big|^{-2} \; C_T(A) \\
    \text{with amplitude amplification } &\rightarrow\quad C_T(U) = \Big| \sin\big( (2j+1) \theta\big) \Big|^{-2} \Big( C_T(A) + j\big(C_T(A)+C_T(A^\dagger)+2C_T(S_\pi)\big) \Big)
\end{align*}
where we reported the expectation value since the RUS construction is non deterministic. For the special case $C_T(S_\pi)\ll C_T(A)$ and $C_T(A)=C_T(A^\dagger)$, amplitude amplification reduces the cost only for $0\leq\lambda_0<\tfrac{1}{3}$, i.e. for relatively low initial success probability. In addition, observe that $C_T(S_\pi)=0$ for $m\leq2$ but grows linearly with $m$ \cite{Jones2013} and therefore cannot be neglected for large $m$.


\section{Controlled RUS operations}
\label{sec:controlled-rus}

It is tempting to consider the RUS construction, after certified success, as fully equivalent to any other coherent quantum operation. This line of reasoning would suggest that RUS constructions can be extended to conditional situations in a naive way. For example, imagine that one desires to implement the quantum phase estimation \cite{Abrams1999} of the operation $U$. Even in its recursive version \cite{Aspuru-Guzik2005}, it is required to condition the action of $U$ with respect to the state of one control qubit being $\ket{1}$. When the control qubit is in state $\ket{0}$, the identity operation is instead performed. This is easily obtained by conditioning all gates forming $A$ on the state of the control qubit: in this case the ancilla qubits for the RUS construction are left in state $\ket{00\dots 0}$, and their measurement always indicates that the identity operation has been successfully performed. Formally, we can define $B$ to be the conditional implementation of $A$:
\begin{equation}
\label{eq:controlled-A}
    B = I^{(m+1)} \otimes \ket{0}\bra{0} + A \otimes \ket{1}\bra{1} \; .
\end{equation}

However, even when immediately successful (meaning that the outcome of the first measurement of the $m$ ancilla qubits is the desired value $00\dots 0$), we show that $B$ distorts the amplitudes of an initial superposition. Here and in the following sections we will discuss ways to reduce such distortion. To clarify the problem, consider the initial state:
\begin{equation}
    \ket{0^m} \ket{\psi} \ket{+} \; ,
\end{equation}
with $\ket{+}=\tfrac{1}{\sqrt 2} (\ket{0}+\ket{1})$ and where we have explicitly separated the three registers ($m$ ancilla qubits, 1 data qubit, and 1 control qubit). Then
\begin{equation}
    B \ket{0^m}\ket{\psi}\ket{+} =
        \tfrac{1}{\sqrt 2} \ket{0^m} \Big(\ket{\psi}\ket{0} + \sqrt{\lambda_0} U \ket{\psi}\ket{1}) \Big)
            + \tfrac{1}{\sqrt 2} \sum_{i=1}^{2^m-1} \sqrt{\lambda_i} \ket{i} R_i \ket{\psi} \ket{1}
\end{equation}
and after measuring the success outcome (i.e. all $m$ ancilla qubits being in state $\ket{0}$) we are left with the two-qubit state
\begin{equation}
\label{eq:distorted-state}
    \frac{1}{\sqrt{1+\lambda_0}} \Big( \ket{\psi}\ket{0} + \sqrt{\lambda_0} U \ket{\psi} \ket{1} \Big) \; ,
\end{equation}
instead of the desired state
$\frac{1}{\sqrt 2} (\ket{\psi}\ket{0} + U \ket{\psi} \ket{1})$.

This effect is what we call ``amplitude distortion'' and it is due to the different success probability of implementing $U$ as compared to the deterministic implementation of the identity operation (having therefore unit success probability). The squared overlap between the state in Eq.~\eqref{eq:distorted-state} and the desired state is $\tfrac{(1+\sqrt{\lambda_0})^2}{2(1+\lambda_0)}$. The situation is even more dramatic in case of measuring an outcome corresponding to failure. In this case, the amplitude corresponding to the control qubit being in $\ket{0}$ is completely erased.

Below we quantify the average overlap considering all possible sequences of failures and success. However, it is intuitively clear that the distortion would be minimized by engineering the same success probability for the identity and for the RUS operations. This can be done in two ways: by reducing the success probability of the identity towards $\lambda_0$ or by increasing that of the RUS construction to 1. Here we present a simple way to achieve the former case, while the latter case will be analyzed in the next sections.

Consider the following modification to $B$:
\begin{align}
    B^\prime &= \Big( D \otimes I \otimes \ket{0}\bra{0} + I^{(m+1)} \otimes \ket{1}\bra{1} \Big) B \nonumber \\
             &= D \otimes I \otimes \ket{0}\bra{0} + A \otimes \ket{1}\bra{1} \; ,
\end{align}
with $D$ being a quantum operation such that
\begin{equation}
    D \ket{0^m} = \sqrt{\gamma_0} \ket{0^m}
                + \sum_{i=1}^{2^m-1} \sqrt{\gamma_i} \ket{i} \; .
\end{equation}
and ideally $\gamma_i\approx\lambda_i$ for $i=0,1,\dots,2^m-1$.

In practice $B^\prime$ leaves the state of the data qubit unchanged compared to $B$, but mimics the action of $A$ on the ancilla register to adjust the probability of measuring any of the $2^m$ outcomes to a value $\gamma_i$ close to $\lambda_i$. However, this approach is likely to have a large cost in terms of $T$ gates, especially if $m>1$ and $\lambda_i$ are not (intuitively speaking and in mathematically vague terms) ``simple'' fractions. If $m=1$, the operation $D$ can be chosen to correspond to a rotation around the y-axis, $D=\exp(-i (\pi/2-\theta) Y)$, with $Y$ being the second Pauli matrix and $\cos(\pi/2-\theta)=\sin(\theta)=\sqrt{\lambda_0}$ as in the previous section.

Since the RUS construction was originally introduced to reduce the decomposition cost of single-qubit rotations, it makes sense to consider this approach only for $D$ having small $T$ cost, i.e. for values of $\lambda_0$ corresponding to easily-decomposable rotations (this is what we meant with ``simple'' in the above paragraph). In addition, the decision of reducing the success probability seems unsatisfactory when there exist alternative techniques that increase the success probability, as we discuss in the next section.

As anticipated above, the rest of the section is devoted to the quantification of the amplitude distortion effect. Consider the action of $B^\prime$ on the most general initial state of the data and control qubits:
$(\alpha\ket{\psi_0}\ket{0} + \beta\ket{\psi_1} \ket{1})$.
The operation we would like to implement is $U$ on the data qubit conditioned on the control qubit being in state $\ket{1}$, otherwise the data qubit is left unchanged when the control qubit is in state $\ket{0}$. Therefore, the desired final state is
$\ket{\varphi} = \alpha \ket{\psi_0}\ket{0} + \beta U \ket{\psi_1} \ket{1}$.
The analysis below includes the implementation of $B$ from eq.~\eqref{eq:controlled-A} as special case by fixing $\gamma_0=1$ and $\gamma_i=0$ for $i>0$.

As detailed in appendix~\ref{app:amplitude-distortion}, the final state realized at the end of a complete RUS procedure depends on the specific sequence of failures before the first success is obtained. It is therefore necessary to average the fidelity with respect to $\ket{\varphi}$ of the states obtained at the end of every acceptable sequence of measurement outcomes weighted by the probability that the sequence is realized. Denoting the average fidelity with $\overline{\mathcal{F}}$, one has:
\begin{align}
\label{eq:avg_fid}
    \overline{\mathcal{F}}
        &= \sum_{k=0}^\infty \; \sum_{i_1,\dots,i_k\neq 0}
            \left( |\alpha|^2 \sqrt{\gamma_0}\,\prod_{j=1}^k \sqrt{\gamma_{i_j}} +
                   |\beta|^2 \sqrt{\lambda_0}\,\prod_{j=1}^k \sqrt{\lambda_{i_j}}
            \right)^2 \nonumber \\
        &= |\alpha|^4
           + 2 |\alpha|^2 |\beta|^2 \frac{\sqrt{\gamma_0\lambda_0}}{1-\sum_{i\neq0} \sqrt{\gamma_i\lambda_i}}
           + |\beta|^4
        \; ,
\end{align}
with $0 \leq \sum_{i \neq 0} \sqrt{\gamma_i\lambda_i} < 1$ when $\gamma_0\lambda_0\neq0$. For the important case $m=1$ that describes, for example, all circuits analyzed in reference \cite{Paetznick2014}, the above expression reduces to
\begin{align}
\label{eq:avg_fid_m1}
    \overline{\mathcal{F}}
        &= \sum_{k=0}^\infty \;
            \left( |\alpha|^2 \sqrt{\gamma_0 (1-\gamma_0 )^k} +
                   |\beta|^2  \sqrt{\lambda_0(1-\lambda_0)^k}
            \right)^2 \nonumber \\
        &= |\alpha|^4 +
            2 |\alpha|^2 |\beta|^2 \sqrt{\gamma_0 \lambda_0}
                \frac{1}{1-\sqrt{(1-\gamma_0)(1-\lambda_0)}} +
            |\beta|^4
        \; .
\end{align}

\begin{figure}[b!]
\centering
\includegraphics[scale=0.38]{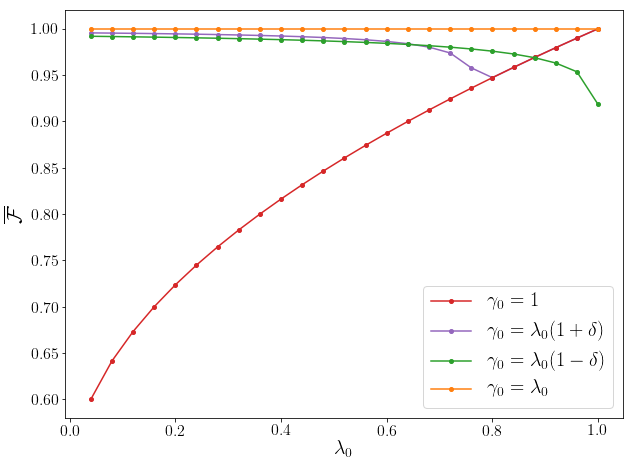}
\quad
\includegraphics[scale=0.38]{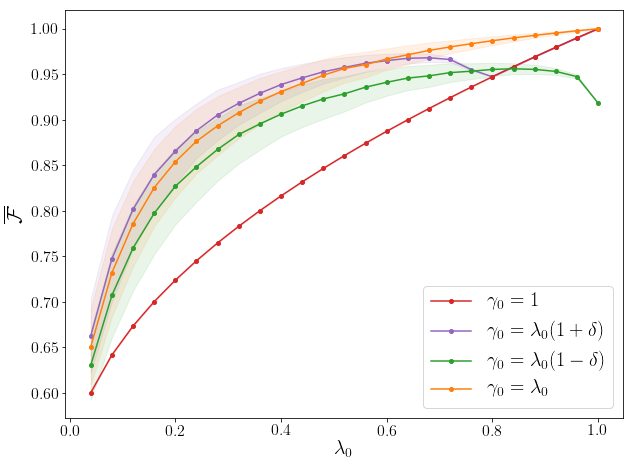}
\caption{Average fidelity $\overline{\mathcal{F}}$ as a function of $\lambda_0$. \textbf{Left panel:} initial amplitudes $\alpha=\beta=1/\sqrt{2}$, number of ancilla qubits for the RUS $m=1$, relative difference between $\lambda_0$ and $\gamma_0$ is $\delta=0.3$. We enforce that $0\leq \gamma_0\leq 1$. For example, this implies that red and purple curves overlap for $\lambda_0\geq 1/(1+\delta)$. \textbf{Right panel:} similar plot with $m=4$ ancilla qubits involved in the RUS construction. While the value of $\gamma_0$ is a deterministic function of $\lambda_0$, for each point we select the remaining $\gamma_i$ and $\lambda_i$ for $i\neq 0$ from a uniform, random distribution. We report the average of 1000 such cases, with the standard deviation indicated by the corresponding shaded areas.}
\label{fig:average-fidelity}
\end{figure}

Fig.~\ref{fig:average-fidelity} provides a graphical visualization of the amplitude distortion in a few relevant situations. The left panel depicts the situation of eq.~\eqref{eq:avg_fid_m1} in which there is a unique measurement outcome $i=1$ associated to the RUS failure: its probability is $\gamma_1 = 1-\gamma_0$ and $\lambda_1 = 1-\lambda_0$ respectively. The four curves represent different relationships between $\gamma_0$ and $\lambda_0$: the red curve is associated with the operator $B$ for which $\gamma_0=1$, while the other curves to operator $B^\prime$ having success probability for the identity operation systematically higher, lower or equal (purple, green, orange lines respectively) than the success probability of the RUS circuit. Here we choose $\delta=0.3$, indicating a 30$\%$ difference between $\gamma_0$ and $\lambda_0$.

The right panel of Fig.~\ref{fig:average-fidelity} shows similar curves for $m=4$, meaning that there are $2^4-1=15$ different outcomes associated with failure. The failure probabilities are stochastically chosen according to a uniform, random distribution for both $\{\gamma_i\}_{i\neq0}$ and $\{\lambda_i\}_{i\neq0}$. The markers correspond to the average over 1000 values of the failure probabilities, and the shaded areas indicate one standard deviation of the corresponding distributions.




\section{Deterministic oblivious amplitude amplification}
\label{sec:deterministic}

In this section and the next, we show how the success probability of the RUS circuits can be increased arbitrarily close to certainty. If the success probability is known, the factor $\lambda_0$ in eq.~\eqref{eq:rus-state} can be amplified to exactly 1 by using variants of oblivious amplitude amplification such that the desired state is obtained with certainty \cite{Brassard2002}. In the context of this work, consider the natural generalization of the reflection $S_\pi$:
\begin{equation}
\label{eq:generalized-reflection}
    S_\phi = \Big(I^m - \big(1-e^{i\phi})\big) \ket{0^m}\bra{0^m}\Big) \otimes I \; .
\end{equation}

Following the reasoning of \cite{Brassard2002}, one can implement $j$ iterations of regular OAA and achieve the state in eq.~\eqref{eq:amplitude-amplification}. Choose $j$ such that $\pi/2 - (2j+1)\theta=\chi$ with $0\leq\chi<2\theta$. In geometrical terms, this means that a further iteration of amplitude amplification would over-rotate the state beyond the desired one. By applying $(-A S_\phi A^\dagger S_\varphi)$ for properly chosen $\phi$ and $\varphi$, the final state is:
\begin{equation}
\label{eq:deterministic-amplitude-amplification}
    (-A S_\phi A^\dagger S_\varphi) (-A S_\pi A^\dagger S_\pi)^j A \ket{0^m}\ket{\psi}
        = \ket{0^m} U \ket{\psi} \; .
\end{equation}
Explicitly, the condition on the phases for the generalized reflections was already provided in reference \cite{Brassard2002}:
\begin{equation}
\label{eq:tan-chi}
    \tan(\chi) = e^{i\varphi} \sin(2\theta)
        \Big( -\cos(2\theta) + i \cot(\phi/2) \Big)^{-1} \; .
\end{equation}

From a practical perspective, this means that from the knowledge of $\lambda_0$ one can directly compute both the number of ``standard'' OAA steps (i.e. those involving the reflection $S_\pi$) and the exact form of the additional, generalized OAA step (uniquely determined by $\phi$ and $\varphi$). Then, by applying the deterministic OAA protocol, one can boost the success probability to certainty.

This approach seems ideal for the unconditional implementation of $U$ and it completely eliminates the amplitude distortion in the conditional case. In terms of T-gate cost, $C_T(S_\phi)$ depends on the value of the phase $\phi$, and we will provide an estimate of its cost, and a comparison with all other methods discussed in this work, in the next section.



\section{Fixed-point oblivious amplitude amplification}
\label{sec:fixed-point}

If the success probability is not known a priori, we propose the application of techniques derived from a variant of the Grover search algorithm \cite{Grover1997} introduced to avoid over-rotation and named $\pi/3$ fixed-point search \cite{Grover2005}. In the context of the RUS circuits, the new protocol is obtained by substituting the original operation $A$ with the sequence $-A S_{\pi/3} A^\dagger S_{\pi/3} A$. Following the reasoning detailed in appendix~\ref{app:fixed-point-oblivious}, the algorithm effectively reduces the probability of failure from its original value $\epsilon=1-\lambda_0$ to the value $\epsilon^3$. Since $\epsilon \leq 1$, the probability of failure is always decreased or, equivalently, the success probability is always increased. The probability of failure is suppressed very effectively when $\lambda_0\approx1$. The state before the ancilla measurement can be explicitly  written as:
\begin{equation}
\label{eq:fixed-point}
    -A S_{\pi/3} A^\dagger S_{\pi/3} A \ket{0^m}\ket{\psi}
        = e^{-2i\pi/3} \sqrt{\lambda_0} \Big( e^{i\pi/3} + 1 - \lambda_0 \Big) \ket{0^m}U\ket{\psi} + e^{-2i\pi/3} (1-\lambda_0)^{3/2} \ket{\Phi^\perp} \; .
\end{equation}

As explained in reference \cite{Grover2005}, the procedure can be concatenated (or nested). To this extent, define
\begin{equation}
\begin{cases}
    A_0 = A \\
    A_k = -A_{k-1} S_{\pi/3} A_{k-1}^\dagger S_{\pi/3} A_{k-1}
\end{cases}
\; ,
\end{equation}
then by applying $A_k$ the error probability is reduced to $\epsilon^{3^k}$. Observe that $A_k$ uses three times more resources than $A_{k-1}$ (if one considers that $A_{k-1}^\dagger$ has a cost similar to $A_{k-1}$ and that the cost of $S_{\pi/3}$ is negligible). Overall, the error probability decreases exponentially in the number of original circuits $A$ used.

While the above algorithm is optimal in a certain sense \cite{Chakraborty2005}, a different fixed-point algorithm has been proposed in reference \cite{Yoder2015} that preserves the quadratic speedup of Grover for $\lambda_0 \ll 1$ and reduces to the $\pi/3$ fixed-search for $\lambda_0\approx 1$. To distinguish it from the previous one, we will call the latter protocol fixed-point oblivious amplitude amplification (FP OAA) as opposed to the $\pi/3$ OAA protocol. Conceptually, the FP OAA protocol corresponds to a sequence of generalized amplitude amplification iterations in which the value of phases $\phi$ and $\varphi$ depends on the length of the protocol. One of the most important properties of FP OOA is that, given the number of iterations $L$, one can provide rigorous performance bounds.

Define $G(\phi,\varphi)=-A^\dagger S_\phi A S_\varphi$, the protocol with length $L$ corresponds to:
\begin{equation}
    A^{(FP)}_L = G(\phi_L,\varphi_L) \dots G(\phi_2,\varphi_2) \, G(\phi_1,\varphi_1) A
\end{equation}
where the subscript ${(FP)}$ is used to distinguished the FP OAA protocol from the $\pi/3$ OAA protocol. By implementing $A^{(FP)}_L$ instead of $A$, the success probability $\lambda_L$ is larger than $(1-\delta)$ if the original success probability $\lambda_0$ was larger than a certain threshold $w$ depending on $L$ and $\delta$. Formally (expressions adapted from \cite{Yoder2015}, notice that the meaning of $L$ and $\delta$ is different in that work):
%
\begin{equation}
\label{eq:validity-fp-oaa}
    \text{given }\delta>0 \quad,\quad
        \lambda_0\geq w=1-\gamma^2
        \,\Rightarrow\, \lambda_L \geq 1-\delta \; ,
\end{equation}
with $\gamma^{-1}=T_{1/(2L+1)}(1/\sqrt{\delta})$ and $T_k(x)=\cos[k\cos^{-1}(x)]$ being the $k^\text{th}$ Chebyshev polynomial of the first kind. For large $L$ and small $\delta$, the threshold $w$ can be approximated as $w\approx (\log(2/\sqrt{\delta})/2L)^2)$ demonstrating that $w$ can be made arbitrary small by increasing the length of the protocol. Suitable values for the phases \cite{Yoder2015} are given by :
\begin{equation}
\label{eq:angles-fp-oaa}
    \phi_j = \varphi_{L-j+1} = - 2\cot^{-1}\Big( \tan\big(2\pi j /(2L+1) \big) \sqrt{1-\gamma^2} \Big) \; .
\end{equation}

Notice that $\phi_j$ and $\varphi_j$ do not depend on $\lambda_0$ and therefore they can be determined from the desired maximum failure probability $\delta$ and a lower bound of the success probability of the original RUS construction $A$ (to play the role of $w$ in the above expressions). The upper bound on the failure probability can be extended to an upper bound on the extent of amplitude distortion for conditional RUS constructions \cite{discussion-with-YC}. However, as discussed in the next section, caution must be exercised to obtained the desired results.

We devote the remaining of this section to, first, quantify the cost of fixed-point OAA techniques in terms of the number of required T gates and, second, estimate the reduction of the amplitude distortion effect for conditional applications of the RUS construction.

\begin{figure}[b]
\centering
\includegraphics[scale=0.35]{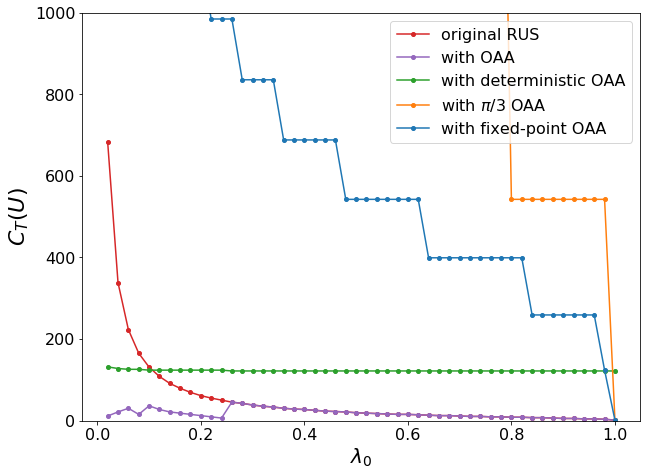}
\quad
\includegraphics[scale=0.35]{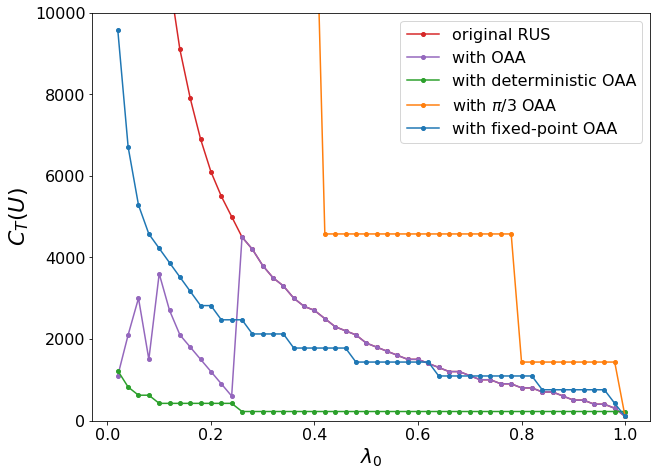}
\caption{T-gate cost $C_T(U)$ as a function of $\lambda_0$. The goal is to achieve an overall success probability $\lambda_0^\text{(fin)}\geq1-10^{-6}$ either through classical repetition or by using OAA protocols. It is important to notice that some amplitude amplification protocols, like standard OAA or deterministic OAA, require the knowledge of $\lambda_0$ and this may not always be the case. \textbf{Left panel:} small cost of the original RUS circuit, namely $C_T(A)=1$. \textbf{Right panel:} large cost of the original RUS circuit, namely $C_T(A)=100$. As a reference, $C_T(S_\phi)\approx60-80$ is a good estimate in the various OAA protocols.}
\label{fig:t-cost}
\end{figure}

Recall that one of the main use of RUS circuits is to reduce the cost in terms of T gates for decomposition of arbitrary rotations using a finite set of gates like $\{H,S,T\}$. It is important to ask what would be the T-gate cost of implementing $\pi/3$ OAA or FP OAA and for what values of $\lambda_0$ one may expect a beneficial reduction. To this extent we focus on the $m=1$ case for which the operator $S_\phi$ in eq.~\eqref{eq:generalized-reflection} corresponds, apart from a global phase, to $R_z(\phi)=\exp(i \phi Z/2)$ with $Z$ being the third Pauli matrix. We quantify the number of T gates required to implement the different OAA protocols, including the fixed-point OAA discussed in this section, to achieve a final success probability $\lambda^{\text{(fin)}}_0 \geq 1-\delta$, with $\delta=10^{-6}$. Our findings are presented in Fig.~\ref{fig:t-cost} and the explicit formulas used to generate the data-points are derived in Appendix~\ref{app:t-cost}.

It is important to notice that the final cost $C_T(U)$ strongly depends on the T-gate cost of a single RUS circuit and of the generalized reflection $S_\phi$, respectively provided as $C_T(A)$ and $C_T(S_\phi)$. The former cost is independent of the OAA protocol used, we choose $C_T(A)=1$ for the left panel and $C_T(A)=100$ for the right panel of Fig.~\ref{fig:t-cost}. The latter cost is harder to determine, especially because a rigorous estimate would require the independent analysis of each OAA protocol under errors in the application of the OAA reflections. We considered $C_T(S_\phi)$ to be equal to the number of T-gates required to decompose $S_\phi$ (not for the specific $\phi$ values in the protocols, but for a generic rotation angle) up to precision $\epsilon=\delta/n_S$, with $n_S$ the number of OAA reflections in the coherent part of the protocol. We refer to appendix~\ref{app:t-cost} for further explanations and for the case $\delta=10^{-3}$.

Finally, it is interesting to study the amplitude distortion effect when the probability of failure is extremely low, as it is the case when fixed-point OAA techniques are applied to RUS circuits. To this scope, we denote with $\lambda^{(FP)}_0$ the success probability amplified using FP OAA, and with $\{\lambda^{(FP)}_i\}_{i\neq0}$ the remaining probabilities of failure.

\begin{figure}[h!]
\centering
\includegraphics[scale=0.4]{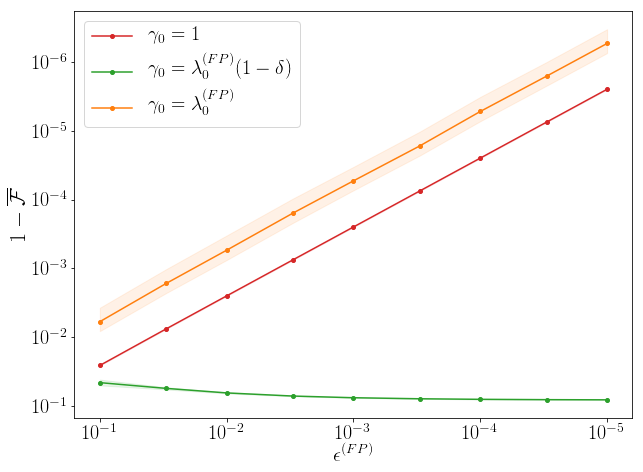}
\caption{Average fidelity $\overline{\mathcal{F}}$ as a function of the probability of failure $\epsilon^{(FP)}$. The plot focuses on the behavior for small $\epsilon$ that can be achieved with the amplitude amplification techniques discussed in section~\ref{sec:fixed-point}. The RUS cisrcuits involves $m=4$ ancilla qubits. While the value of $\gamma_0$ is a deterministic function of $\lambda^{(FP)}_0$, for each point we select the remaining $\gamma_i$ and $\lambda^{(FP)}_i$ for $i\neq 0$ from a uniform, random distribution. We report the average of 1000 such cases, with the standard deviation indicated by the corresponding shaded areas. For the green curve, $\delta=0.3$.}
\label{fig:average-fidelity-2}
\end{figure}

Fig.~\ref{fig:average-fidelity-2} depicts the situation of eq.~\eqref{eq:avg_fid} with $m=4$ ancilla qubits involved in the RUS circuit. The three curves represent different relationships between $\gamma_0$ and $\lambda^{(FP)}_0$: the red curve is associated with the operator $B$ for which $\gamma_0=1$, while the other curves to operator $B^\prime$ having success probability for the identity operation systematically lower or equal (green and orange lines respectively) than the success probability of the RUS circuit. Notice that, in both $B$ and $B^\prime$, the original circuit $A$ is effectively substituted with its OAA version. Here we choose $\delta=0.3$, indicating a 30$\%$ difference between $\gamma_0$ and $\lambda^{(FP)}_0$ and, in practice, $\gamma_0\approx 0.7$.

Since $m=4$, there are 15 different outcomes associated with failure. The corresponding probabilities are stochastically chosen according to a uniform, random distribution for both $\{\gamma_i\}_{i\neq0}$ and $\{\lambda^{(FP)}_i\}_{i\neq0}$. By construction, $\sum_{i\neq 0} \lambda^{(FP)}_i = \epsilon^{(FP)}$. The markers in Fig.~\ref{fig:average-fidelity-2} correspond to the average over 1000 values of the failure probabilities, and the shaded areas indicate one standard deviation of the corresponding distributions. Notice that the figure is, in practice, a magnification of the rightmost side of Fig.~\ref{fig:average-fidelity} (right panel) if one was to substitute $\epsilon^{(FP)}=1-\lambda_0$.

\section{Current limitations of conditional RUS circuits}
\label{sec:limitations}

Amplitude distortion in the implementation of conditional RUS operations is due to the non-unit success probability $\lambda_0$. Considering the operation $B$ defined in eq.~\eqref{eq:controlled-A} and the initial state $\ket{\psi}\otimes(\alpha\ket{0}+\beta\ket{1})$, the absolute values of the superposition amplitudes $\alpha$ and $\beta$ are preserved (1) in an exact way using deterministic OAA if $\lambda_0$ is known a priori, (2) in an approximated way with bounded error using fixed-point OAA, and (3) in an approximated way using $\pi/3$ OAA if $\lambda_0$ is unknown.

Observe that both $\pi/3$ and fixed-point OAA protocols introduce a global phase associated to the desired state $\ket{\Phi}=\ket{0^m}U\ket{\psi}$ that depends on $\lambda_0$. This phase is usually neglected in the protocol characterization since one focuses on the success probability alone. However, the ``global'' phase becomes a relative phase in the context of conditional implementation of $U$ through a RUS construction and may give rise to problems if not properly taken into account.

In case $\lambda_0$ is known, the unwanted phase can be corrected by single-qubit rotations (around the z-axis) on the control qubit. If $\lambda_0$ is unknown, one can only rely on the preservation of the absolute value of the $\beta$ amplitude. This result is unsatisfactory, but improved protocols may solve or mitigate this effect.
Nevertheless, we can envision three situation in which the relative phase can be corrected or is largely uninfluential.

First, when one has to implement $B$ followed by a sequence of gates that are (block) diagonal in the computational basis of the control qubit and terminating with $B^\dagger$. In fact it is possible to introduce an opposite phase by using the $\pi/3$ protocol with $S_{-\pi/3}$ instead of $S_{\pi/3}$ (see appendix~\ref{app:fixed-point-oblivious}) based on a RUS circuit with the same probability of success $\lambda_0$. Notice that the probability of success of $A^\dagger$ is the same as that of $A$ (see appendix~\ref{app:inverse-rus}).

Second, one can use the $\pi/3$ and $-\pi/3$ protocols in an alternating way to implement even powers of $U$ in a conditional way without spurious phases.

Third, certain algorithms rely solely on the absolute value of the amplitudes (corresponding to $\alpha$ and $\beta$ above, i.e. considering the computational basis for the state of the control qubit). As a concrete example, the recently proposed design of a quantum neuron \cite{Cao2017a} is based on conditional RUS operations. Even with unmitigated amplitude distortion, networks based on multiple quantum neurons show the ability to learn boolean functions (in the context of binary classifiers) and exhibit attractor dynamics along the lines of associative memories. Using fixed-search protocols is expected to improve the design and provide rigorous upper bounds for the number of required RUS repetitions.

Finally, we observe that the T-gate cost of the unconditional protocol is not a direct proxy of the cost of the conditional protocol: For example, a controlled CNOT results in a Toffoli gate that requires T gates to be decomposed in the $\{H,S,T,CNOT\}$ universal gate set. We reserve to analyze these implications in future works.


\section{Conclusions}

Repeat-Until-Success (RUS) circuits are coherent, non-deterministic quantum operations. Their analysis had so far neglected situations in which their execution is controlled by the state of one or more qubits, but our study reveals that the non-zero probability of failure of the single RUS attempt gives rise to a distortion of the quantum amplitudes of the output state. We propose two ways for mitigating this effect, both based on the observation that state transitions for the RUS ancilla qubits should be the same whether or not the RUS circuit is actually implemented. The first approach leaves unaltered the success probability $\lambda_0$ of the RUS circuit, but introduces additional operations for the ancilla qubits, whereas the second method relies on techniques that boost the original success probability towards certainty.

Our contribution is twofold: on one hand we clarify the origin and quantify the relevance of amplitude distortion in conditional implementations of RUS circuits. On the other hand we adapt fixed-point oblivious amplitude amplification (OAA) to the RUS context and apply it to cases in which the value of $\lambda_0$ is not known a priori (however in the FP OAA case a lower bound must be known). Since RUS circuits are mainly used to reduced the T-gate count of quantum operations, we provide a comparison of the T-gate cost of several (both standard and fixed-point) OAA protocols.

Our results suggest that fixed-point OAA may be more expensive than the original RUS circuit, especially when its T-gate cost is very small to begin with and the success probability already relatively good. However, fixed-point OAA cancels the amplitude distortion effect and, therefore, becomes a forced choice in conditional settings when $\lambda_0$ is a priori unknown. For example, we expect that fixed-point OAA may play an important role in future quantum neural networks proposals \cite{discussion-with-YC} along the lines of reference \cite{Cao2017a}.


\begin{acknowledgments}
The author would like to thank Aram Harrow for discussions on fixed-point quantum search techniques and Nicolas Sawaya and Yudong Cao for their helpful feedback on the manuscript.
\end{acknowledgments}


\appendix
\renewcommand\thefigure{\thesection.\arabic{figure}}
\renewcommand\thetable{\thesection.\arabic{table}}
\setcounter{figure}{0}
\setcounter{table}{0}

\bigskip\noindent\makebox[\linewidth]{\resizebox{0.3333\linewidth}{1pt}{$\bullet$}}\bigskip


\section{Quantification of the amplitude distortion effect}
\label{app:amplitude-distortion}

Consider the conditional implementation of the RUS circuit. According to section~\ref{sec:controlled-rus} in the main text, we can describe the effect of $B^\prime$ as:
\begin{align}
    B^\prime \ket{0^m}(\alpha \ket{\psi_0}\ket{0} + \beta \ket{\psi_1}\ket{1})
            =& \ket{0^m} \Big(\alpha \sqrt{\gamma_0} \ket{\psi_0} \ket{0} +
                            \beta \sqrt{\lambda_0} U \ket{\psi_1}\ket{1} \Big) +\nonumber \\
            &\quad \sum_{i=1}^{2^m-1} \ket{i} \Big(\alpha \sqrt{\gamma_i} \ket{\psi_0} \ket{0} +
                                            \beta \sqrt{\lambda_i} R_i \ket{\psi_1} \ket{1} \Big) \; ,
\end{align}
where $\gamma_i$ should ideally be close in value to $\lambda_i$. If $\gamma_i=\lambda_i$ for every $i=0,1,2,\dots ,2^m-1$, then no distortion is generated. Otherwise the amplitude $\alpha$ and $\beta$ of the desired output
$\ket{\varphi}=(\alpha\ket{\psi_0}\ket{0}+\beta U \ket{\psi_1}\ket{1})$
are modified.

The amplitude distortion depends on the sequence of measurement outcomes before success is observed. We aim at quantifying the average distortion of the conditional RUS operation based on $B^\prime$.

Consider the situation in which one has $k$ failures before achieving success, meaning that the measurement outcomes of the ancillas corresponds to a certain sequence
$i_1, i_2, \dots , i_k, i_{k+1}$ such that $i_{k+1}=0$ and $i_j\neq 0$ for $j\leq k$.
The state (of the data and control qubits) after the $k$ failures and corrections is given by
\begin{equation}
    \ket{\varphi_k(i_1,i_2,\dots,i_k)} = \left(
        \alpha \bigg( \prod_{j=1}^k \sqrt{ \gamma_{i_j}} \bigg) \ket{\psi_0}\ket{0}+
        \beta  \bigg( \prod_{j=1}^k \sqrt{\lambda_{i_j}} \bigg) \ket{\psi_1}\ket{1}
        \right) \bigg/ \sqrt{n_k(i_1,i_2,\dots,i_k)}
        \; .
\end{equation}
where the normalization factor is
\begin{equation}
    n_k(i_1,i_2,\dots,i_k) = |\alpha|^2 \prod_{j=1}^k \gamma_{i_j} +
                                |\beta|^2 \prod_{j=1}^k \lambda_{i_j}\; .
\end{equation}

Considering state $\ket{\varphi_{k-1}(i_1,i_2,\dots,i_{k-1})}$ and the action of $B^\prime$, the conditional probability of observing the outcome $i_k$ given the previous outcomes $i_1,\dots,i_{k-1}$ is:
\begin{align}
    p_k(i_k|i_1,i_2,\dots,i_{k-1})
    &= \left( |\alpha|^2 \,\prod_{j=1}^k \gamma_{i_j} +
           |\beta|^2 \,\prod_{j=1}^k \lambda_{i_j} \right)
        \bigg/
        n_{k-1}(i_1,i_2,\dots,i_{k-1}) \nonumber \\
    &=  n_k(i_1,i_2,\dots,i_k) \,
        \big/ \, n_{k-1}(i_1,i_2,\dots,i_{k-1})
    \; .
\end{align}

Denoting with $p_k(i_1,i_2,\dots,i_k)$ the probability of the corresponding sequence of outcomes, from the recursive relations:
\begin{equation}
\begin{cases}
    p_1(i_1) = |\alpha|^2 \gamma_{i_1} + |\beta|^2 \lambda_{i_1} = n_1(i_1) \\
    p_k(i_1,i_2,\dots,i_k) = p_{k-1}(i_1,i_2,\dots,i_{k-1}) \; p_k(i_k|i_1,i_2,\dots,i_{k-1})
\end{cases}
\end{equation}
one obtains the explicit expression:
\begin{equation}
    p_k(i_1,i_2,\dots,i_k)  = n_k(i_1,i_2,\dots,i_k)
                            =|\alpha|^2 \,\prod_{j=1}^k \gamma_{i_j} +
                                |\beta|^2 \,\prod_{j=1}^k \lambda_{i_j}\; .
\end{equation}

Finally, considering also the successful outcome at the $(k+1)$ attempt, the state of the data and control qubits is
\begin{equation}
    \ket{\varphi_{k+1}(i_1,i_2,\dots,i_k,0)} = \left(
        \alpha \bigg( \sqrt{ \gamma_0}\,\prod_{j=1}^k \sqrt{ \gamma_{i_j}} \bigg) \ket{\psi_0}\ket{0}+
        \beta  \bigg( \sqrt{\lambda_0}\,\prod_{j=1}^k \sqrt{\lambda_{i_j}} \bigg) U\ket{\psi_1}\ket{1}
        \right) \bigg/ \sqrt{p_{k+1}(i_1,i_2,\dots,i_k,0)}
        \; .
\end{equation}

Due to the stochasticity of the measurement procedure, the average state at the end of the conditional RUS operation is given by the incoherent sum of the different sequences of failures and success weighted by their probability:
\begin{equation}
    \rho = \sum_{k=0}^\infty \; \sum_{i_1,\dots,i_k\neq 0}
                p_{k+1} \ket{\varphi_{k+1}} \bra{\varphi_{k+1}}\; ,
\end{equation}
where we have omitted the dependency from $(i_1,i_2,\dots,i_k,0)$. Since one must obtain the first success at a certain point, then it is clear that
$\sum_{k=0}^\infty \; \sum_{i_1,\dots,i_k\neq 0} p_{k+1}(i_1,i_2,\dots,i_k,0)=1$
. The average (squared) overlap with the desired state is then:
\begin{align}
    \overline{\mathcal{F}}
        &\equiv \bra{\varphi}\rho\ket{\varphi} \\
        &= \sum_{k=0}^\infty \; \sum_{i_1,\dots,i_k\neq 0} p_{k+1}(i_1,i_2,\dots,i_k,0) \,
            \left| \braket{\varphi | \varphi_{k+1}(i_1,i_2,\dots,i_k,0)} \right|^2 \nonumber \\
        &= \sum_{k=0}^\infty \; \sum_{i_1,\dots,i_k\neq 0} p_{k+1}
            \left| \frac{ |\alpha|^2 \sqrt{\gamma_0}\,\prod_{j=1}^k \sqrt{\gamma_{i_j}} +
                          |\beta|^2 \sqrt{\lambda_0}\,\prod_{j=1}^k \sqrt{\lambda_{i_j}}}
                        {\sqrt{p_{k+1}}} \right|^2 \nonumber \\
        &= \sum_{k=0}^\infty \; \sum_{i_1,\dots,i_k\neq 0}
            \left( |\alpha|^2 \sqrt{\gamma_0}\,\prod_{j=1}^k \sqrt{\gamma_{i_j}} +
                   |\beta|^2 \sqrt{\lambda_0}\,\prod_{j=1}^k \sqrt{\lambda_{i_j}}
                   \right)^2 \nonumber \\
        &= |\alpha|^4
           + 2 |\alpha|^2 |\beta|^2 \frac{\sqrt{\gamma_0\lambda_0}}{\Gamma}
           + |\beta|^4
        \; ,
\end{align}
with $\Gamma = 1 - \sum_{i \neq 0} \sqrt{\gamma_i\lambda_i}$. Since $\{\gamma_i\}_i$ and $\{\lambda_i\}_i$ are probability distributions, it follows that $0\leq \Gamma \leq 1$.


\vspace{0.3cm}
\section{Proof of $\pi/3$ fixed-point oblivious amplitude amplification}
\label{app:fixed-point-oblivious}

Consider eq.~\eqref{eq:rus-state} in the main text. By introducing the following two $(m+1)$-qubit states, namely $\ket{\Psi}=\ket{0^m}\ket{\psi}$ and $\ket{\Phi}=\ket{0^m}U\ket{\psi}$, the expression can be rewritten as:
\begin{equation}
    A \ket{\Psi} = \sqrt{\lambda_0} \ket{\Phi}
            + \sqrt{1-\lambda_0} \ket{\Phi^\perp} \; ,
\end{equation}
where the success probability is given by $\lambda_0$ and the $(m+1)$-qubit state $\ket{\Phi^\perp}$ has no overlap with the support of projector $\Pi=\ket{0^m}\bra{0^m}\otimes I$. Define the state $\ket{\Psi^\perp}$ according to the equation:
\begin{equation}
    A \ket{\Psi^\perp} =\sqrt{1-\lambda_0} \ket{\Phi} - \sqrt{\lambda_0} \ket{\Phi^\perp} \; .
\end{equation}
State $\ket{\Psi^\perp}$ is orthogonal to $\ket{\Psi}$ and the ``2D Subspace Lemma'' of reference \cite{Berry2014} shows that it also has no overlap with the support of $\Pi$. Then one obtains
\begin{equation}
    \Pi A^\dagger \ket{\Phi}
        = \sqrt{\lambda_0} \Pi \ket{\Psi} + \sqrt{1-\lambda_0} \Pi \ket{\Psi^\perp}
        = \sqrt{\lambda_0} \ket{\Psi} \; .
\end{equation}

Finally, recall the generalized reflection $S_\phi$ defined in eq.~\eqref{eq:generalized-reflection} and denote with $I$ the identity operator on all $(m+1)$ qubits. Using $(1-e^{i\pi/3})=e^{-i\pi/3}$ and $(-1+e^{-i\pi/3})=e^{-2i\pi/3}$, one can show that
\begin{align}
    (-A S_{\pi/3} A^\dagger S_{\pi/3}) A \ket{\Psi}
        &= - A \Big( I - (1-e^{i\pi/3})\Pi \Big)  A^\dagger \Big( I - (1-e^{i\pi/3})\Pi \Big) A \ket{\Psi} \nonumber \\
        &= -A \Big( I -e^{-i\pi/3}  A^\dagger \Pi A - e^{-i\pi/3} \Pi + e^{-i2\pi/3} \Pi A^\dagger \Pi A \Big) \ket{\Psi}  \nonumber \\
        &= -A \ket{\Psi} + e^{-i\pi/3} \Pi A \ket{\Psi} + e^{-i\pi/3} A \Pi \ket{\Psi} - e^{-2i\pi/3} A\Pi A^\dagger \Pi A \ket{\Psi} \nonumber \\
        &= -A \ket{\Psi} + e^{-i\pi/3} \sqrt{\lambda_0} \ket{\Phi} + e^{-i\pi/3} A \ket{\Psi} - e^{-2i\pi/3} \sqrt{\lambda_0} A\Pi A^\dagger \ket{\Phi} \nonumber \\
        &= \Big( -1 + e^{-i\pi/3} - e^{-2i\pi/3} \lambda_0 \Big) A \ket{\Psi} + e^{-i\pi/3} \sqrt{\lambda_0} \ket{\Phi} \nonumber \\
        &= e^{-2i\pi/3} (1 - \lambda_0) \Big( \sqrt{\lambda_0} \ket{\Phi} + \sqrt{1-\lambda_0} \ket{\Phi^\perp} \Big) + e^{-i\pi/3} \sqrt{\lambda_0} \ket{\Phi} \nonumber \\
        &= e^{-2i\pi/3} \sqrt{\lambda_0} \Big( e^{i\pi/3} + 1 - \lambda_0 \Big) \ket{\Phi} + e^{-2i\pi/3} (1-\lambda_0)^{3/2} \ket{\Phi^\perp} \; .
\end{align}
Denoting $\lambda_0 = 1-\epsilon$, one can see that the probability of failure of the original RUS circuit, \emph{i.e.} $\epsilon$, is reduced to $\epsilon^3$ by applying $(-A S_{\pi/3} A^\dagger S_{\pi/3} A)$ instead of $A$ alone.

Finally, observe the effect of having phase $-\pi/3$ in the generalized reflections. The explicit expression reported below clarifies that the amplitudes for both $\ket{\Phi}$ and $\ket{\Phi^\perp}$ have opposite phase compared to $(-A S_{\pi/3} A^\dagger S_{\pi/3}) A \ket{\Psi}$:
\begin{equation}
    (-A S_{-\pi/3} A^\dagger S_{-\pi/3}) A \ket{\Psi}
        = e^{2i\pi/3} \sqrt{\lambda_0} \Big( e^{-i\pi/3} + 1 - \lambda_0 \Big) \ket{\Phi} + e^{2i\pi/3} (1-\lambda_0)^{3/2} \ket{\Phi^\perp} \; .
\end{equation}


\vspace{0.5cm}
\section{RUS construction for the inverse operation}
\label{app:inverse-rus}

Recall eq.~\eqref{eq:rus-state} in the main text and the definitions of appendix~\ref{app:fixed-point-oblivious}, namely the initial state $\ket{\Psi}=\ket{0^m}\ket{\psi}$, the desired state $\ket{\Phi}=\ket{0^m}U\ket{\psi}$ and the two states $\ket{\Phi^\perp}$ and $\ket{\Psi^\perp}$ outside the support of $\Pi=\ket{0^m}\bra{0^m}\otimes I$. They are related by
\begin{equation}
\begin{cases}
    A \ket{\Psi}
            = \sqrt{\lambda_0} \ket{\Phi} + \sqrt{1-\lambda_0} \ket{\Phi^\perp} \\
    A \ket{\Psi^\perp}
            = \sqrt{1-\lambda_0} \ket{\Phi} - \sqrt{\lambda_0} \ket{\Phi^\perp}
\end{cases}
\; .
\end{equation}
or equivalently by
\begin{equation}
\begin{cases}
    A^\dagger \ket{\Phi}
        = \sqrt{\lambda_0} \ket{\Psi} + \sqrt{1-\lambda_0} \ket{\Psi^\perp} \\
    A^\dagger \ket{\Phi^\perp}
        = \sqrt{1-\lambda_0} \ket{\Psi} - \sqrt{\lambda_0} \ket{\Psi^\perp}
\end{cases}
\; .
\end{equation}

Observe that the expressions are valid for every $m$-qubit state $\ket{\psi}$ of the date register and that, denoting $\ket{\psi}=U^\dagger\ket{\phi}$, the symmetry of the above expressions makes it clear that
\begin{center}
\hspace{4mm}if $A\,\;$ is the RUS construction of $U\,\;$ with success probability $\lambda_0$, \\
then $A^\dagger$ is the RUS construction of $U^\dagger$ with success probability $\lambda_0$.
\end{center}


\vspace{0.5cm}
\section{T-gate cost for RUS circuits with various OAA protocols}
\label{app:t-cost}

The data-points in Fig.~\ref{fig:t-cost} of the main text are obtained according to the expressions in this appendix section. For each approach, the scope is boosting the original success probability $\lambda_0$ above the threshold $(1-\delta)$. It is considered that $C_T(A^{-1})=C_T(A)$ and that, by limiting our analysis to the case $m=1$, $C_T(S_\pi)=0$ while in general $C_T(S_\phi)$ is expected to be similar to $C_T(R_z(\phi))$ (effectively the same operation up to a global phase when $m=1$).

The original RUS formulation only requires classical repetitions of $A$. For an overall success probability $\lambda_0^\text{(fin)} \geq 1-\delta$ one needs at least $k$ repetitions, where $k$ is the smallest integer such that
\begin{equation*}
    \sum_{j=0}^{k} (1-\lambda_0)^j\lambda_0 \geq 1-\delta \; ,
\end{equation*}
resulting in $k = \lceil \log(\delta)/\log(1-\lambda_0) -1 \rceil$ and a corresponding T-gate cost of
\begin{equation}
    \text{without OAA } \rightarrow\quad\
        C_T(U) = C_T(A) \lceil \log(\delta)/\log(1-\lambda_0) -1 \rceil \; .
\end{equation}

Applying the oblivious amplitude amplification protocol may not suffice to reach the desired success probability without multiple classical repetitions since the error of the protocol is not guaranteed to go below $\mathcal{O}(\lambda_0)$. Formally, the last step before over-rotation is associated with success probability $\lambda^{\text{OAA}}_0 = \left| \cos(\chi) \right|^2$, where the angle $\chi$ already appeared in section~\ref{sec:deterministic} in the context of deterministic OAA.
The T-gate cost for an overall (meaning OAA + classical repetitions) success probability above the threshold $(1-\delta)$ is
\begin{equation}
    \text{with OAA } \rightarrow\quad\
        C_T(U) = (2j+1) \, C_T(A) \, \lceil \log(\delta)/\log(\sin^2\chi) -1 \rceil \; .
\end{equation}
with $\chi=\pi/2 - (2j+1)\theta$ determined by $\theta=\arcsin(\sqrt\lambda_0)$ and $j=\lfloor (\pi/2\theta -1)/2 \rfloor$.

In alternative to classical repetitions of the OAA protocol, an additional (generalized) iteration can be performed to achieve deterministic OAA. The T-gate cost is:
\begin{equation}
    \text{deterministic OAA } \rightarrow\quad\
        C_T(U) = 2(j+1) \, C_T(A) + C_T(S_\phi) + C_T(S_\varphi) \; .
\end{equation}
where angles $\phi$ and $\varphi$ are determined by eq.~\eqref{eq:tan-chi} of the main text, reported below for convenience:
\begin{equation*}
    \tan(\chi) = e^{i\varphi} \sin(2\theta) \Big( -\cos(2\theta) + i \cot(\phi/2) \Big)^{-1} \; .
\end{equation*}
Observe that there is no dependence on the threshold $(1-\delta)$ since the protocol is deterministic and has unitary success probability.

The success probability can also be amplified by using fixed-point OAA protocols. For the $\pi/3$ OAA protocol discussed in section~\ref{sec:fixed-point} of the main text, recall that each level of concatenation reduces the error doubly exponentially, but also cost exponentially. Let us compute the cost explicitly. The $k$-level concatenation has failure probability $(1-\lambda_0)^{3^k}$ and therefore the desired success probability is achieved when
\begin{equation*}
    (1-\lambda_0)^{(3^k)} \leq \delta \quad\implies\quad
        k = \Big\lceil \frac{\log\log(1/\delta)-\log\log (1/(1-\lambda_0))}{\log 3} \Big\rceil \; .
\end{equation*}
The corresponding cost is determined by the recurrence relation $C_T(A_j) = 3 C_T(A_{j-1}) + 2 C_T(S_{\pi/3})$ that can be expressed as the order-2 homogeneous linear recurrence with constant coefficients $C_T(A_j) = 4 C_T(A_{j-1})-3 C_T(A_{j-2})$. The associated characteristic polynomial is $p(t)=t^2-4t+3=(t-3)(t-1)$ that leads to the closed formula $C_T(A_j) = a 3^j + b$. The constants $a, b$ are determined by the first two values of the sequence, in our case: $C_T(A_0)=C_T(A)$ and $C_T(A_1)=3C_T(A)+2C_T(S_{\pi/3})$.
The T-gate cost for $k$-level concatenation is then:
\begin{equation}
    \text{$\pi/3$ OAA } \rightarrow\quad\
        C_T(U) = \Big( C_T(A) + C_T(S_{\pi/3}) \Big) \, 3^k - C_T(S_{\pi/3}) \; .
\end{equation}
where $k$ is explicitly provided in the equation above.

Finally, let us consider the fixed-point OAA protocol. From eq.\eqref{eq:validity-fp-oaa} and the other definitions in section~\ref{sec:fixed-point}, one can first determine the length $L$ of the protocol by choosing, for the specific $\lambda_0$ of the RUS under consideration, the smallest integer $L$ such that
\begin{equation*}
    \lambda_0 \geq  1 -  \left[ T_{1/(2L+1)}(1/\sqrt{\delta}) \right]^{-2} \; ,
\end{equation*}
with $T_j(x)=\cos[j \cos^{-1}(x)]$ being the $j^\text{th}$ Chebyshev polynomial of the first kind. The T-gate cost then results:
\begin{equation}
    \text{fixed-point OAA } \rightarrow\quad\
        C_T(U) = (2L+1) C_T(A) + \sum_{j=1}^L \left( C_T(S_{\phi_j}) + C_T(S_{\varphi_j}) \right) \; ,
\end{equation}
with $\phi_j$ and $\varphi_j$ given by eq.~\eqref{eq:angles-fp-oaa} in the main text.

It is important to add two remarks: first, the cost $C_T(S_\phi) \approx C_T(R_z(\phi))$ is not computed for the specific angle $\phi$ but using the average T-gate cost solely determined by the precision $\epsilon$ that is requested on the operation $R_z(\phi)$. Second, it is unclear what value of $\epsilon$ should be required for the various OAA protocols so that the result of the protocol is effectively $U$.

However, without studying the error stability of the various OAA protocols, it is intuitively reasonable to require that $S_\phi$ are decomposed within precision given by the threshold $\delta$ divided by the number of generalized reflections needed in the coherent part of each protocol. Indicating the required precision with $\epsilon$, we then consider $C_T(S_\phi)=3.21 \log_2(1/\epsilon)-6.93$ as indicated by the scaling of the KMM method \cite{Kliuchnikov2016} as reported in reference \cite{Paetznick2014}.

\begin{figure}[h!]
\centering
\includegraphics[scale=0.3]{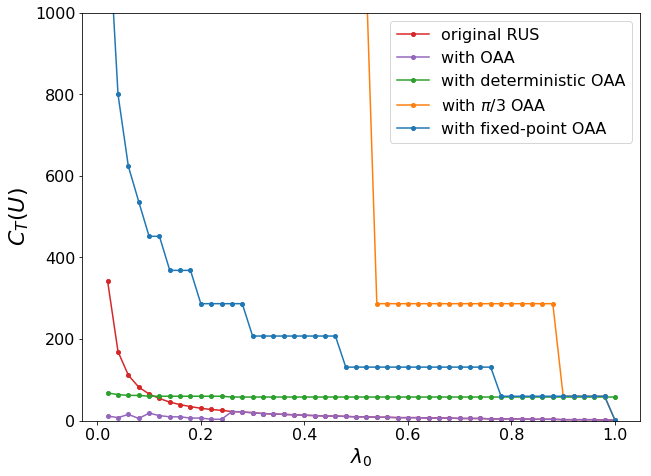}
\quad
\includegraphics[scale=0.3]{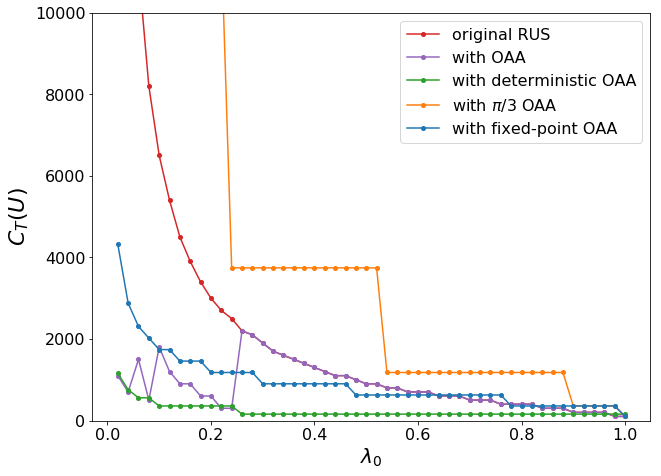}
\caption{T-gate cost $C_T(U)$ as a function of $\lambda_0$. The goal is to achieve an amplified success probability $\lambda_)^\text{(fin)}\geq1-10^{-3}$. \textbf{Left panel:} small cost of the original RUS circuit, namely $C_T(A)=1$. \textbf{Right panel:} large cost of the original RUS circuit, namely $C_T(A)=100$. As a reference, $C_T(S_\phi)\approx30-50$ is a good estimate in the various OAA protocols.}
\label{app:fig:t-cost}
\end{figure}

In the main text we reported the T-gate cost to achieve an amplified success probability above the threshold $(1-\delta)$, with $\delta=10^{-6}$. In Fig.~\ref{app:fig:t-cost} we consider the case $\delta=10^{-3}$.


\bibliographystyle{unsrt}
\bibliography{references}

\end{document}